\documentclass[aps,twocolumn,a4paper,prl,superscriptaddress,floatfix,showpacs,nofootinbib]{revtex4}

\usepackage{amssymb}
\usepackage{amsmath}
\usepackage{dsfont}
\usepackage{amsfonts}
\usepackage{graphicx}
\usepackage{graphics}
\usepackage{color}

\def\TE{\mathrm{TE}}
\def\TM{\mathrm{TM}}

\def\dd{\mathrm{d}}

\def\P{\mathrm{p}}
\def\B{\mathrm{B}}

\def\poln{\mathrm{l}}
\def\polx{\mathrm{x}}
\def\polm{\mathrm{m}}

\def\perf{\mathrm{P}}
\def\Drud{\mathrm{D}}
\def\DP{\mathrm{D,P}}
\def\NT{0}
\def\HT{\mathrm{HT}}

\def\max{\mathrm{max}}
\def\lmax{\ell_\max}

\def\txi{\tilde{\xi}}
\def\tsigma{\tilde{\sigma}}
\def\tomega{\tilde{\omega}}

\def\calF{\mathcal{F}}
\def\calE{\mathcal{E}}
\def\calS{\mathcal{S}}
\def\calR{\mathcal{R}}
\def\calD{\mathcal{D}}
\def\calI{\mathcal{I}}
\def\calK{\mathcal{K}}
\def\calL{\mathcal{L}}

\begin{document}

\title{Classical Casimir interaction in the plane-sphere geometry}

\author{Antoine Canaguier-Durand}
\altaffiliation{Present address: ISIS - Universit\'{e} de Strasbourg.}
\affiliation{Laboratoire Kastler Brossel, ENS, UPMC, CNRS, F-75252
Paris, France}
\author{Gert-Ludwig Ingold}
\affiliation{Institut f{\"u}r Physik, Universit{\"a}t Augsburg,
D-86135 Augsburg, Germany}
\author{Marc-Thierry Jaekel}
\affiliation{Laboratoire de Physique Th{\'e}orique, ENS, UPMC, CNRS,
F-75231 Paris, France}
\author{Astrid Lambrecht}
\affiliation{Laboratoire Kastler Brossel, ENS, UPMC, CNRS, F-75252
Paris, France}
\author{Paulo A. Maia Neto}
\affiliation{Instituto de F{\'\i}sica, UFRJ, CP 68528, Rio de
Janeiro, RJ, 21941-909, Brazil}
\author{Serge Reynaud}
\affiliation{Laboratoire Kastler Brossel, ENS, UPMC, CNRS, F-75252
Paris, France}
\date{\today }

\begin{abstract}
We study the Casimir interaction in the plane-sphere geometry in the
classical limit of high temperatures. In this limit, the finite
conductivity of the metallic plates needs to be taken into account.
For the Drude model, the classical Casimir interaction is
nevertheless found to be independent of the conductivity so that it
can be described by a single universal function depending only on
the aspect ratio $x=L/R$ where $L$ is the interplate distance and
$R$ the sphere radius. This universal function differs from the one
found for perfect reflectors and is in principle amenable to
experimental tests. The asymptotic approach of the exact result to
the Proximity Force Approximation appears to be well fitted by polynomial expansions in $\ln
x$.
\end{abstract}


\maketitle

The Casimir effect arises due to the confinement of the quantum
fluctuations of electromagnetic fields between reflecting bodies.
Its discussion is often focused on the ideal case introduced by H.
Casimir \cite{Casimir} of a pair of perfectly reflecting parallel
plates at zero temperature. In this case, the energy is given by a
universal expression $\calE^\perf_\NT=-\hbar c \pi ^2 A/720L^3 $
depending only on geometrical parameters, the mirrors' separation
$L$ and surface $A$, and fundamental constants, $\hbar$ and $c$.
However this idealization does not hold for any experiment and
reliable descriptions of the Casimir interaction have to account for
the optical response of matter \cite{Lifshitz56,Jaekel91}.

In the present letter, we want to study another case where the
Casimir interaction becomes universal, namely the high-temperature
limit indicated in the following by a subscript $\HT$. This limit
has been thoroughly studied for two perfectly reflecting parallel
plates \cite{Mehra67,Brown69,Schwinger78} where the free energy
$\calF^\perf_\HT =-\zeta(3) k_\B T A/8\pi L^2$ depends only on
temperature $T$ and geometrical parameters $L$ and $A$ ($\zeta$ is
the Riemann function and $\zeta(3)\simeq1.202$; the superscript
$\perf$ refers to perfect mirrors). When the Drude model is used for
describing the effect of conduction electrons in metallic mirrors,
the free energy $\calF^\Drud_\HT =\calF^\perf_\HT/2$ is one half of
that obtained for perfect mirrors
\cite{Bostrom00,BrevikNJP06,IngoldPRE09} (the superscript $\Drud$
refers to the Drude model).

This reduction by a factor 2 at large temperatures, independent of
the parameters of the Drude model (see below), is confirmed by
microscopic descriptions of the interaction between two metallic
bulks \cite{Jancovici05,Buenzli05,Bimonte09}. This difference can in
principle be tested at room temperature by measuring the force at
large distances. Experiments are difficult in this domain because
the force decreases with distance. However, experiments have
recently been performed at distances up to 7$\mu$m where the thermal
effect is large. The results have been interpreted by the authors
\cite{Sushkov11} as being in agreement with the Drude model
prediction once an electrostatic patch contribution is subtracted
\cite{Behunin12}. This conclusion stands in contradiction with the
results of other Casimir force measurements, performed at smaller
distances, up to 0.75$\mu$m, which were interpreted by their authors
as excluding the dissipative Drude model and agreeing with the
lossless plasma model \cite{Decca0507}. This contradiction remains a
matter of debate (see discussions and references in
\cite{KlimRMP09,LambrechtCasimir11}).

It is also worth stressing that most precise experiments are
performed in the plane-sphere geometry. The evaluation of the force
is often done through the so-called \textit{Proximity Force
Approximation} (PFA) \cite{Derjaguin68} which amounts to averaging
the force between parallel plates over the distribution of local
inter-plate distances. This trivial treatment of geometry cannot
reproduce the rich relation expected between Casimir effect and
geometry \cite{Balian,WeberPRD10,RahiCasimir11}. Theoretical
treatments going beyond the PFA have recently been proposed, in
particular through a multipolar expansion well adapted to the
plane-sphere geometry. In the following, we will use results known
for perfect or metallic mirrors coupled to electromagnetic fields
\cite{Maia08Canaguier09,Canaguier10}.

The aim of the present letter is to investigate the classical
Casimir interaction for the geometry of a plane and a sphere with
arbitrary aspect ratio. We will focus our attention on the models of
perfect mirrors and dissipative mirrors and use scattering
methods discussed in \cite{Canaguier10}. The results will turn out
to be given by universal functions of the aspect ratio, for dissipative as
well as perfect mirrors, with the two functions differing from each
other. The limit of high temperatures will allow us to reach a much
better numerical accuracy than in previous works \cite{Canaguier10}.
This improvement in the accuracy will be used to sharpen the
discussion of the asymptotic approach of the exact result to the
PFA. This asymptotic behavior will appear to be well fitted by
polynomial expansions of the logarithm of the aspect ratio.


We start from the Matsubara formula giving the Casimir free energy
$\calF$ between a plane and a sphere \cite{Canaguier10}
\begin{eqnarray}
\label{scattering_formula}
&&\calF  = k_\B T \sum_n^\prime \ln \det \calD  (i\xi_n) ~ , \nonumber  \\
&&\calD  = \calI - \calR _2 e^{-\calK\calL} \calR _1 e^{-\calK\calL}
~ .
\end{eqnarray}
Here $\xi_n=2\pi n k_\B T/\hbar$ are the Matsubara frequencies, and
the prime indicates a sum over integers $n = 0 \ldots\infty $ with a
factor $\frac{1}{2}$ for $n=0$; $\calR _1$ and $\calR _2$ are the
reflection operators on the plane and the sphere respectively,
discussed below in terms of Fresnel reflection amplitudes and Mie
scattering amplitudes; $e^{-\calK \calL }$ describes the one-way
propagation on the distance $\calL=L+R$ between the center of the
sphere and the plane with $R$ the radius of the sphere and $L$ the
minimal distance between the plane and the sphere; $\calK
=\sqrt{k^2+\xi^2/c^2} $ is the longitudinal wavevector after a Wick
rotation from real to imaginary frequencies, with $k$ the modulus of
the transverse wavevector.

The nonzero Matsubara frequencies $\xi_{n\neq0}\ge 2\pi k_\B
T/\hbar$ increase with temperature and their contributions to the
sum (\ref{scattering_formula}) become exponentially small because of
the propagation factor $e^{-\calK \calL }$ with $\calK \ge\xi/c\ge
2\pi k_\B T/\hbar c$. At the high temperature limit $k_\B T \gg
\hbar c /\calL$, the free energy is dominated by the first term
$n=0$ in the sum (\ref{scattering_formula}) and, therefore, it is
proportional to temperature
\begin{eqnarray}
\label{scattering_HT} &&\calF_\HT  = -k_\B T \Phi  \;,\quad \Phi
= -\frac12 \ln \det \calD (0) ~ , \nonumber\\
&&\calD(0) = \calI - \calR_S(0) e^{-k\calL} \calR_P(0) e^{-k\calL} ~
.
\end{eqnarray}
$\Phi $ is a dimensionless function of the parameters which does not
depend on temperature. Hence the entropy $\calS$ is independent of
temperature and the energy $\calE$ vanishes
\begin{eqnarray}
\label{FhighT} &&\calS_\HT =-\partial_T\calF= k_\B \Phi ~ , \\
&& \calE_\HT =\calF_\HT+T\calS_\HT =0  ~ . \nonumber
\end{eqnarray}
This implies that the classical limit of the Casimir interaction has
a purely entropic origin \cite{Feinberg01,Spruch02}.

The Drude model is the simplest description of a dissipative mirror. It corresponds to a dielectric function $\varepsilon
\left[i\xi\right] = 1 + \sigma/\xi$ at imaginary frequencies
$\omega=i\xi$, without a magnetic response ($\mu=1$). $\sigma
\left[i\xi\right] = \omega_\P^2/(\gamma+\xi)$ measures the
conductivity associated with conduction electrons (the SI
conductivity is $\epsilon_0\sigma$). The Drude model is
characterized by two dimensional parameters, the squared plasma
frequency $\omega_\P^2$ proportional to the density of the
conduction electrons, and their relaxation rate $\gamma$. It meets
the important property of ordinary metals to have a finite static
conductivity $\sigma_0 = \omega_\P^2/\gamma$. The lossless plasma
model $\gamma \to 0$ corresponds to an infinite value for
$\sigma_0$, in contradiction with tabulated optical data at low
frequencies \cite{LambrechtEPJ00,SvetovoyPRB08}.

The Fresnel reflection amplitudes $r_\TE$ and $r_\TM$ for plane
waves with polarization TE or TM are well known from studies of the
plane-plane geometry \cite{Bostrom00,BrevikNJP06,IngoldPRE09}. They
are $r_\TE^\perf \to -1$ and $r_\TM^\perf \to 1$ for perfect mirrors
whereas they go to $r_\TE^\Drud \to 0$ and $r_\TM^\Drud \to 1$ for
the Drude model at low frequencies. For the lossless plasma model,
they are found to interpolate between perfect and Drude models, a
property found also below for the Mie amplitudes. The disappearance
of the TE reflection amplitude is responsible for the ratio 2
between the classical limits of the free energies obtained from
perfect and Drude models in the plane-plane geometry.

A similar discrepancy arises also for the Mie amplitudes $a_\ell$
and $b_\ell$ which describe the scattering on the sphere
\cite{Canaguier10}. For perfect mirrors and low frequencies, we get
$a_\ell^\perf \simeq \frac{(-1)^\ell (\ell+1)
\tilde{\xi}^{(2\ell+1)}}{\ell(2\ell+1)!! (2\ell-1)!!}$ and
$b_\ell^\perf \simeq - \frac{a_\ell^\perf \ell}{\ell+1}$. The
electric and magnetic multipoles $a_\ell$ and $b_\ell$ progressively
decrease with increasing multipole index $\ell$, but they have the
same power law in the reduced frequency $\txi=\xi R/c$ for a given
$\ell$. The situation is different for the Drude model where
$a_\ell^\Drud \simeq a_\ell^\perf$ and $b_\ell^\Drud \simeq
\frac{b_\ell^\perf \tsigma_0 \txi }{(2\ell+3)(2\ell+1)}$
($\tsigma_0= \sigma_0 R/c$ is the reduced frequency associated with
the static conductivity). Incidentally, the results for the lossless
plasma model are found to interpolate between perfect and Drude
models: $b_\ell$ has the same form as $b_\ell^\perf$ (resp.
$b_\ell^\Drud$) for large spheres (resp. small spheres). The
crossover between these two regimes is determined by the parameter
$\tomega_\P$ being respectively larger or smaller than 1
($\tomega_\P=\omega_\P R/c$ is the reduced plasma frequency).

The function $\Phi^\perf$ is a universal expression of the aspect
ratio for perfect mirrors, as there exist no other parameters. The
discussion of the preceding paragraph entails that it is also the
case for the function $\Phi^\Drud$ calculated for Drude mirrors \cite{Zandi2010},
although the latter is associated with dimensional parameters like
$\sigma_0$. As a matter of fact, $b_\ell$ is negligible with respect
to $a_\ell$ for any given value of $\ell$, and it follows that the
classical high temperature limit is the same as if $b_\ell^\Drud$
would simply be zero. In spite of the fact that $\Phi^\Drud$ and
$\Phi^\perf$ are universal functions of the aspect ratio, these two
functions differ. As another important consequence of this
discussion, the lossless plasma model result does not obey the
universality property met by perfect as well as Drude mirrors. For
this reason, we will not discuss this model any longer in this
letter.

In the remainder of this letter, we present the results of the
numerical evaluation of the functions $\Phi^\DP$ which determine the
classical Casimir free energy between a plane and a sphere, for
Drude and perfect mirrors respectively. We write these functions
through a comparison with the often used PFA approximation (the
superscript $\DP$ means $\Drud$ or $\perf$)
\begin{eqnarray}
\label{defrho} &&\Phi ^\DP (x)= \frac{C^\DP}{x} \varrho^\DP(x)
\;,\quad x\equiv \frac LR ~ , \nonumber\\
&&C^\Drud\equiv \frac{\zeta(3)}{8} \;,\quad C^\perf\equiv
\frac{\zeta(3)}{4}~ .
\end{eqnarray}
The first factors in the expressions of $\Phi^\DP$ are the PFA
results while the second ones, $\varrho^\DP$, represent deviations
from PFA. The factors $\varrho^\DP$ go to 1 for large spheres
$x\to0$, which confirms the validity of PFA as an asymptotic
approximation. The fact that $\Phi^\DP$ and $\varrho^\DP$ depend on
a single geometrical parameter, the aspect ratio $x$, corresponds to
the universality discussed above.

\begin{figure}[tbh]
\centering
\includegraphics[width=0.35\textwidth]{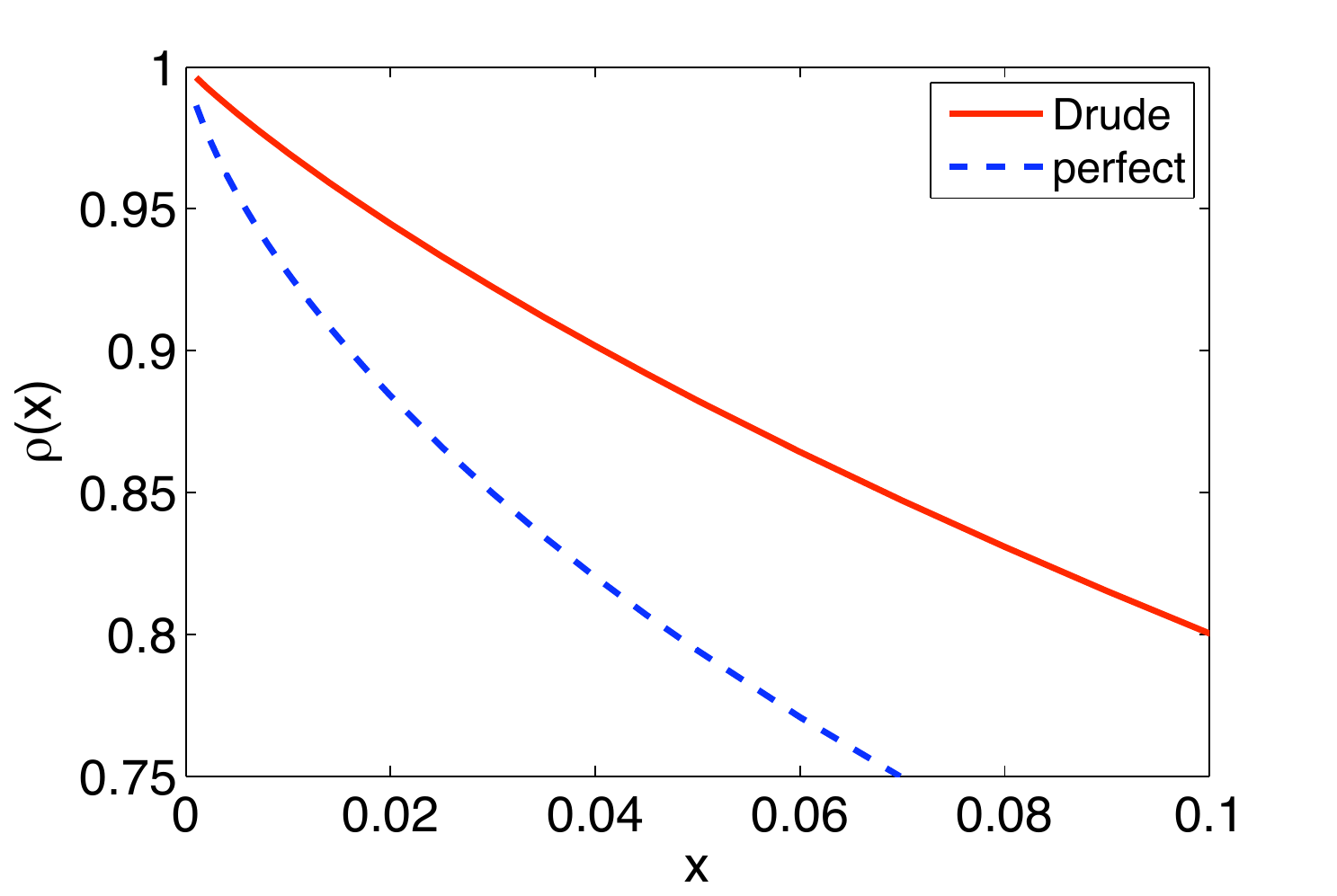} %
\caption{Classical (high temperature) Casimir interaction between
plane and spherical mirrors, represented by the ratio $\varrho^\DP$
of exact result to PFA result versus the aspect ratio $x=L/R$; the
full red line corresponds to Drude mirrors and the dashed blue line
to perfect mirrors [Colors online]. } \label{figure:rhoDP}
\end{figure}

The explicit numerical evaluation proceeds as in \cite{Canaguier10}.
Due to the relatively simple form of the matrices in the low
frequency limit, the numerics can be pushed farther than for the
general case. Precisely, the computation can be pushed to a larger
maximum value $\lmax$ for the multipole index $\ell$, and it follows
that the numerical results are now more accurate for small values of
the aspect ratio $x$. The results of the numerical evaluations shown
on Fig.~\ref{figure:rhoDP} have been calculated with $\lmax$ up to
5000 at our minimal value for the aspect ratio $10^{-3}$. The
numerical accuracy $\delta\varrho$ is estimated to be of the order
of $10^{-4}$ at $x=10^{-3}$ and of $10^{-8}$ at $x=2\times10^{-3}$.

Though both of them are universal, the functions $\varrho^\Drud(x)$
and $\varrho^\perf(x)$ drawn on Fig.~\ref{figure:rhoDP} do not
coincide. In other words, the ratio $\Phi ^\perf(x)/\Phi ^\Drud(x) =
2 \varrho^\perf(x)/\varrho^\Drud(x)$ between Casimir free energies
calculated for perfect and Drude mirrors is itself a function of
$x$. This ratio reaches the value 2 at the PFA limit ($x\to0$), as
known from the studies on plane-plane geometry  \cite{Bostrom00},
and the value $3/2$ at large distances ($x\to\infty$)
\cite{Canaguier10}.

We finally discuss the asymptotic approach to PFA for small values
of $x$. To this aim, we introduce another representation of the
deviation from PFA
\begin{equation}
\varrho^\DP(x)\equiv 1+x\beta^\DP(x) ~ . \label{defbeta}
\end{equation}
$\beta^\DP(x)$ is the slope of the line which joins the points
($0,1$) to ($x,\varrho^\DP(x)$) on the plots of
Fig.~\ref{figure:rhoDP}. It may also be thought of as an additive
correction to PFA in (\ref{defrho})
\begin{eqnarray}
\label{addiPhi} &&\Phi^\DP (x)= C^\DP \left( \frac1x+\beta^\DP(x)
\right) ~ .
\end{eqnarray}
The numerical accuracy $\delta\varrho$ discussed above is now
translated into a numerical accuracy $\delta\beta=\delta\varrho/x$
for $\beta$. This accuracy $\delta\beta$ is estimated to be of the
order of $10^{-1}$ at $x=10^{-3}$ and of $5\times10^{-6}$ at
$x=2\times10^{-3}$. The following discussions of the asymptotic
approach to PFA are possible only thanks to this good accuracy.

\begin{figure}[tbh]
\centering
\includegraphics[width=0.35\textwidth]{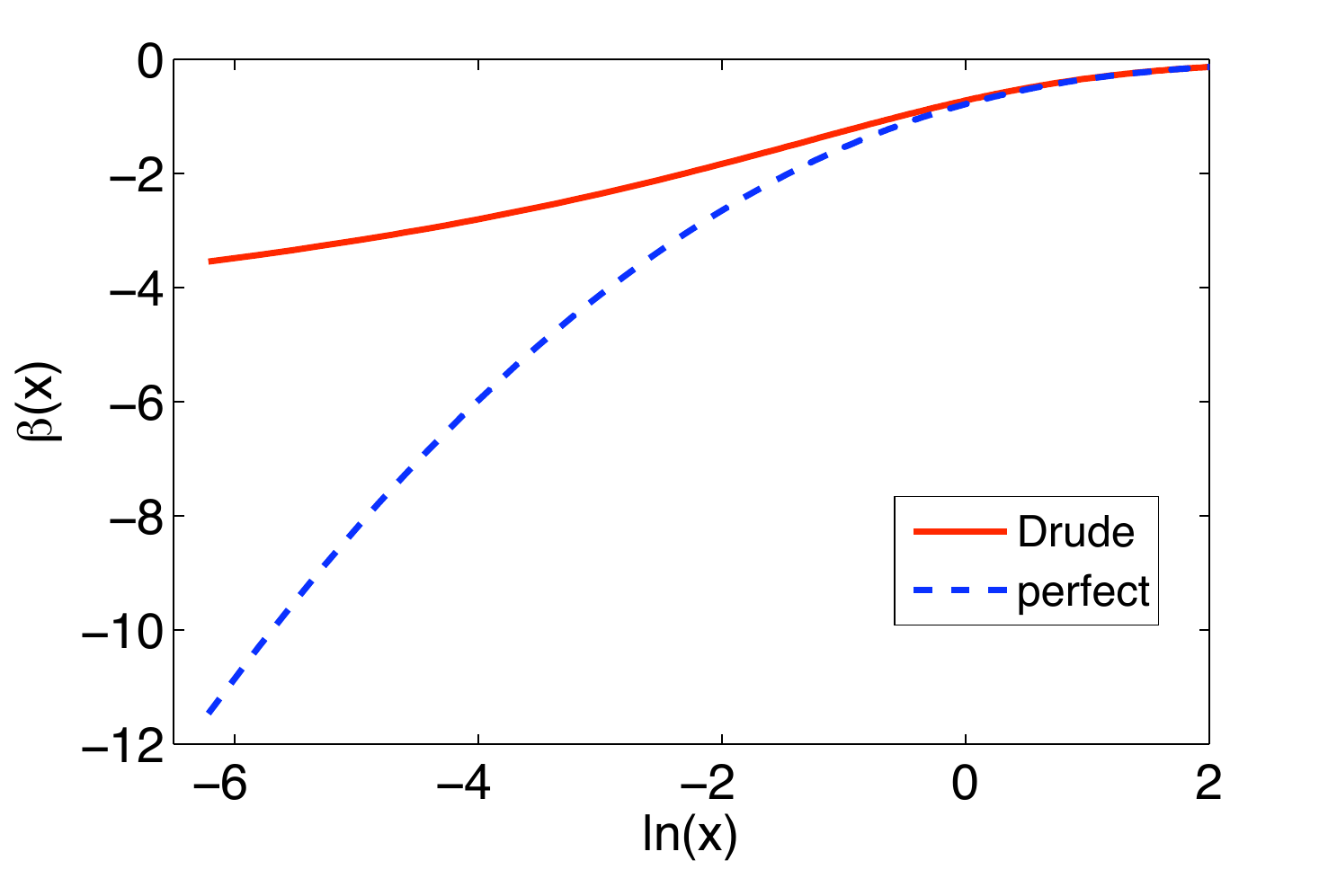} %
 \includegraphics[width=0.357\textwidth]{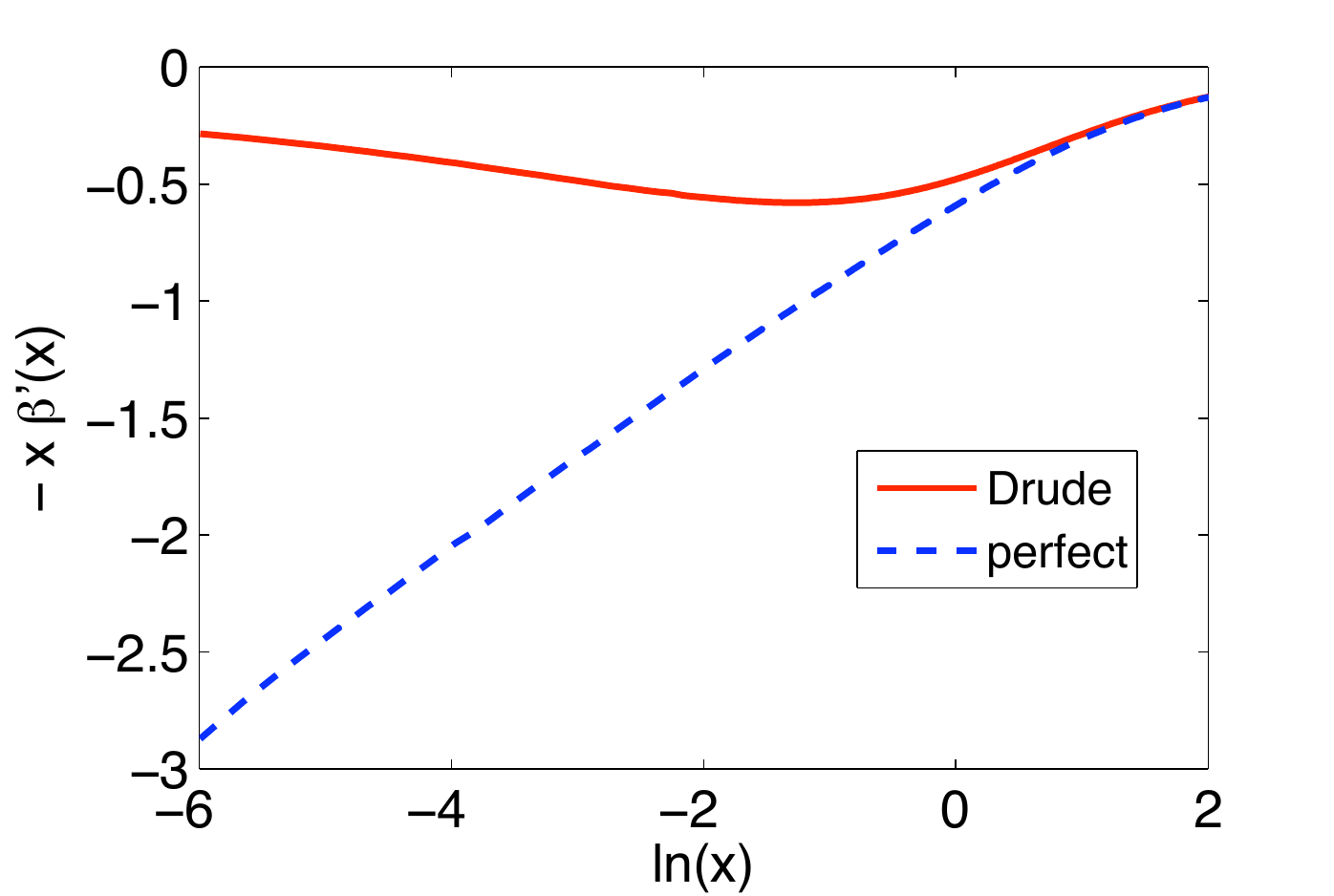} \hphantom{.}%
\caption{Deviation from PFA of the classical Casimir interaction
between plane and spherical mirrors, described by the additive
correction $\beta^\DP(x)$ versus $\ln x$ (upper plot) and the
logarithmic derivatives $-\frac{\dd\beta^\DP}{\dd\ln
x}=-x\frac{\dd\beta^\DP}{\dd x}$ (lower plot); the full red lines
correspond to Drude mirrors and the dashed blue lines to perfect
mirrors [Colors online].} \label{figure:betaDP}
\end{figure}

The quantities $\beta^\DP$ are shown as the upper plot on
Fig.~\ref{figure:betaDP} for Drude and perfect mirrors, as functions
of $\ln x$. The lower plot on Fig.~\ref{figure:betaDP} represents
the quantities $-\frac{\dd\beta^\DP}{\dd\ln x}$ which appear in the
Casimir force
\begin{eqnarray}
\label{addiFor} &&F^\DP (x)= -\frac{C^\DP k_\B T}{L} \left( \frac1x
- \frac{\dd\beta^\DP(x)}{\dd\ln x} \right) ~ .
\end{eqnarray}

The plots on Fig.~\ref{figure:betaDP} suggest that the asymptotic
approach to PFA is well described for small values of $x$ by a
polynomial expansion in the variable $\ln x$. In order to check this
idea, we have performed best fits of the numerical results with the
following trial functions, defined with the same number of
parameters,
\begin{eqnarray}
\label{trialf} &&\beta_\poln (x)= a_0 + a_1 \ln x + a_2 \ln^2 x
+ a_3 \ln^3 x ~ , \nonumber \\
&&\beta_\polx (x)= b_0 + b_1 x + b_2 x^2 + b_3 x^3 ~ , \nonumber \\
&&\beta_\polm (x)=c_0 + c_1 \ln x + c_2 x + c_3 x^2~ .
\end{eqnarray}
The first one, $\beta_\poln$, is a polynomial form in the variable
$\ln x$ as suggested by Fig.~\ref{figure:betaDP} (see also
\cite{BordagPRD09}). The second one, $\beta_\polx$, is a polynomial
form in the variable $x$, which was sufficient to represent
functions $\rho$ obtained from older calculations with a lesser
accuracy \cite{Maia08Canaguier09,Emig08}. It corresponds to the free
energy being a Laurent series in the variable $x$. The third one,
$\beta_\polm$, is a mixed form which results, in the classical limit
(high temperatures), from the assumption \cite{Bimonte11} that the
force (not the free energy) may be written as a Laurent series in
$x$.

\begin{figure}[tbh]
\includegraphics[width=0.38\textwidth]{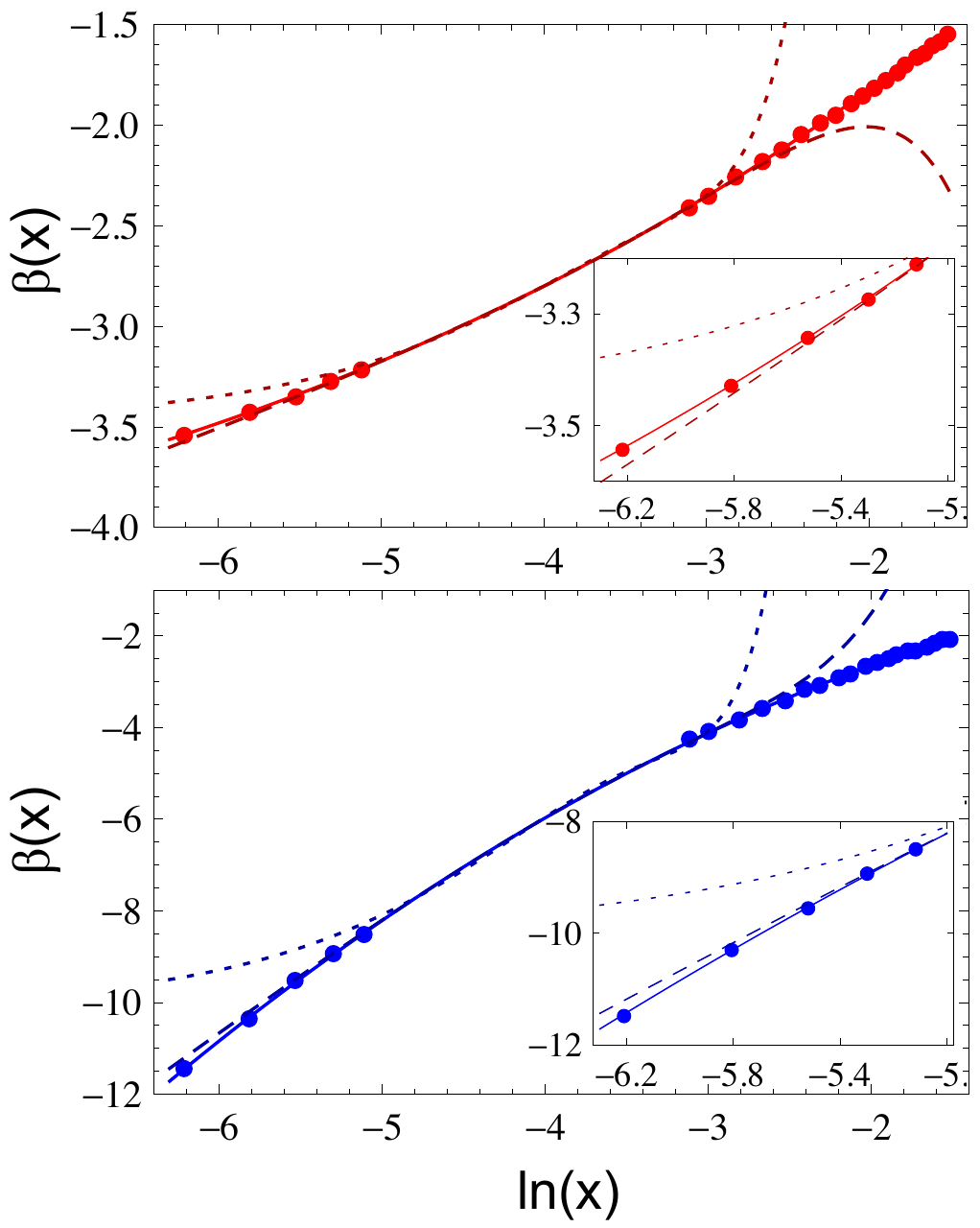} %
\caption{Best fits of the numerical results with the trial functions
(\ref{trialf}) for the Drude (red lines, upper plot) and perfect
mirrors (blue lines, lower plot); the full, dotted and dashed lines
correspond respectively to $\beta_\poln$, $\beta_\polx$ and
$\beta_\polm$; the full circles correspond to numerical points not
used for the fits; the insets zoom on the smallest values of $x$
[Colors online]. } \label{figure:fitsDP}
\end{figure}

In each case, the best fit functions are defined from the numerical
results obtained in the window $\ln x \in \left[ -5 , -3 \right]$.
Their comparison on Fig.~\ref{figure:fitsDP} with the numerical
values obtained in a broader window shows that $\beta_\poln$ is a
better representation of the numerical results than $\beta_\polm$,
itself better than $\beta_\polx$. In particular, the insets on
Fig.~\ref{figure:fitsDP} show the deviations of the best fits of
$\beta_\polx$ and $\beta_\polm$ from the exact results at the
smallest values of the aspect ratio $\ln x \in \left[ -6.2 , -5
\right]$. The assumptions that the free energy or force be a Laurent
series in $x$ cannot be considered as generally valid.

The difference between the cases of Drude and perfect 
 mirrors is in principle amenable to experimental tests. Direct comparison with
exact calculations would require measurements at large distances
(high temperature limit). This challenge could be met with experimental
parameters not so far from those  in the experiment \cite{Sushkov11}.

\acknowledgments The authors thank the ESF Research Networking
Programme CASIMIR (www.casimir-network.com) for providing excellent
possibilities for discussions and exchange. Financial support from
international programs Capes-Cofecub and Procope is gratefully
acknowledged.


\begin{thebibliography}{99}
\newcommand{\REVIEW}[4]{\textrm{#1} \textbf{#2}, #3, (#4)}
\newcommand{\NAME}[4]{#1 #2.#3.#4.}
\newcommand{\Review}[1]{\textrm{#1}}
\newcommand{\Volume}[1]{\textbf{#1}}
\newcommand{\Book}[1]{\textit{#1}}
\newcommand{\Eprint}[1]{\textsf{#1}}
\def\etal{\textit{et al }}


\bibitem{Casimir}
Casimir H. B. G., {Proc. K. Ned. Akad. Wet.} \Volume{51}, 793
(1948)

\bibitem{Lifshitz56}
Lifshitz E. M.,
Sov. Phys. JETP \Volume{2}, 73 (1956)

\bibitem{Jaekel91}
Jaekel M.-T., and Reynaud S.,
J. Physique I \Volume{1}, 1395 (1991)

\bibitem{Mehra67}
Mehra J.,
Physica \Volume{37}, 145 (1967)

\bibitem{Brown69}
Brown L. S., and Maclay G. J.,
Phys. Rev. \Volume{184}, 1272 (1969)

\bibitem{Schwinger78}
Schwinger J., de Raad L. L., and Milton K. A.,
Ann. Phys. NY \Volume{115}, 1 (1978)

\bibitem{Bostrom00}
Bostr\"{o}m M., and Sernelius B. E.,
Phys. Rev. Lett. \Volume{84}, 4757 (2000)

\bibitem{BrevikNJP06}
Brevik I., Ellingsen S. E., and Milton K.A.,
New J. Phys. \Volume{8}, 236 (2006)

\bibitem{IngoldPRE09}
Ingold G.-L., Lambrecht A., and Reynaud S.,
Phys. Rev. E \Volume{80}, 041113 (2009)

\bibitem{Jancovici05}
Jancovici B., and {\v S}amaj L.,
Europhys. Lett. \Volume{72}, 35 (2005)

\bibitem{Buenzli05}
Buenzli P. R., and Martin P. A.,
Europhys. Lett. \Volume{72}, 42 (2005)

\bibitem{Bimonte09}
Bimonte G.,
Phys. Rev. A \Volume{79}, 042107 (2009)

\bibitem{Sushkov11}
Sushkov A. O., Kim W. J., Dalvit D. A. R. and Lamoreaux S.K., Nat.
Phys. \Volume{7}, 230 (2011).

\bibitem{Behunin12}
Behunin R. O., Intravaia F., Dalvit D. A. R., Maia Neto P. A. and
Reynaud S.,
Phys. Rev. A \Volume{85}, 012504  (2012)

\bibitem{Decca0507}
Decca R. S., L\'{o}pez D., Fischbach E., Klimchitskaya, G. L.,
Krause, D. E., and Mostepanenko, V. M., Ann. Phys. NY \Volume{318},
37 (2005);
Phys. Rev. D \Volume{75}, 077101 (2007)

\bibitem{KlimRMP09}
Klimchitskaya G. L., Mohideen U. and Mostepanenko V. M., Rev. Mod.
Phys. \Volume{81}, 1827 (2009)

\bibitem{LambrechtCasimir11}
Lambrecht A., Canaguier-Durand A., Gu\'erout R. and Reynaud S., in
\textit{Casimir physics}, D.A.R. Dalvit \etal eds, Lecture Notes in
Physics 834 (Springer-Verlag, 2011) p.97

\bibitem{Derjaguin68}
Derjaguin B. V., Abrikosova I. I., and Lifshitz E. M., Q. Rev. Chem.
Soc. \Volume{10},  295 (1968)

\bibitem{Balian}
Balian R., and Duplantier B.,
Ann. Phys. NY \Volume{104}, 300 (1977);
Ann. Phys. NY \Volume{112}, 165 (1978)

\bibitem{WeberPRD10}
Weber A., and Gies H.,
Phys. Rev. D \Volume{82}, 125019 (2010)

\bibitem{Zandi2010}
Zandi R., Emig T., and Mohideen U.,
Phys. Rev. B \Volume{81}, 195423 (2010)

\bibitem{RahiCasimir11}
Rahi S.J., Emig T. and Jaffe R.L., in \textit{Casimir physics},
D.A.R. Dalvit \etal eds, Lecture Notes in Physics 834
(Springer-Verlag, 2011) p.129

\bibitem{Maia08Canaguier09}
Maia Neto P. A., Lambrecht A., and Reynaud S.,
Phys. Rev. A \Volume{78}, 012115 (2008); Canaguier-Durand A., Maia
Neto P. A., Cavero-Pelaez I., Lambrecht A., and Reynaud S.,
Phys. Rev. Lett. \Volume{102}, 230404 (2009)

\bibitem{Canaguier10}
Canaguier-Durand A., Maia Neto P. A., Lambrecht A., and Reynaud S.,
Phys. Rev. Lett. \Volume{104}, 040403 (2010);
Phys. Rev. A \Volume{82}, 012511 (2010)

\bibitem{Feinberg01}
Feinberg J., Mann A. and Revzen M., Ann. Phys. NY \Volume{288} 103
(2001)

\bibitem{Spruch02}
Spruch L., Phys. Rev. A \Volume{66}, 022103 (2002)

\bibitem{LambrechtEPJ00}
Lambrecht A., and Reynaud S.,
Eur. Phys. J. D \Volume{8}, 309 (2000)

\bibitem{SvetovoyPRB08}
Svetovoy V. B., van Zwol P. J., Palasantzas G., and De Hosson J. T.
M.,
Phys. Rev. B \Volume{77}, 035439 (2008)

\bibitem{BordagPRD09}
Bordag M., and Nikolaev V.,
Phys. Rev. D \Volume{81}, 065011 (2010)

\bibitem{Emig08}
Emig T.,
J. Stat. Mech. Theory Exp., P04007 (2008)

\bibitem{Bimonte11}
Bimonte G., Emig T., Jaffe R.L. and Kardar M.,
EPL \Volume{97}, 50001 (2012)

\end{thebibliography}
\end{document}